# Evolving hypernetwork model based on WeChat user relations[*]

Fu-Hong Wang  Jin-Li Guo[+]  Ai-Zhong Shen  Qi Suo

*Business School, University of Shanghai for Science and Technology,*

*Shanghai, 200093 ,China*

**Abstract:** Based on the theory of hypernetwork and WeChat online social relations, the paper proposes an evolving hypernetwork model with the competitiveness and the age of nodes. In the model, nodes arrive at the system in accordance with Poisson process and are gradual aging. We analyze the model by using a Poisson process theory and a continuous technique, and give a characteristic equation of hyperdegrees. We obtain the stationary average hyperdegree distribution of the hypernetwork by the characteristic equation. The numerical simulations of the models agree with the analytical results well. It is expected that our work may give help to the study of WeChat information transmission dynamics and mobile e-commerce.

[*] Project supported by the National Natural Science Foundation of China (Grant No. 71571119).

[+] Corresponding author. E-mail: phd5816@163.com

# 微信用户关系超网络演化模型

王福红　郭进利　沈爱忠　索琪

（上海理工大学 管理学院，上海 200093）

**摘要**：本文基于超网络理论，以微信在线社交关系为背景，建立节点具有吸引力与老化的非均齐微信社交关系超网络模型。依据实际微信社交关系的特点，该模型中节点按照 Poisson 过程连续时间进入网络，并且随着时间的增长和社交平台的更新，节点不断老化。我们获得模型超度的特征方程，利用特征方程对模型进行解析，计算机模拟结果与理论分析的结果一致。本文结论为微商用户的营销管理提供理论上借鉴和参考。

**关键词**：社交关系；超网络； 节点老化； 演化模型

# 1 引言

随着互联网和信息技术的发展，出现了许多在线社交工具，如 facebook、微信、QQ 等等，在线社交工具的发展使人类的社交关系从现实的物理空间转向虚拟的在线网络空间。传统的社交网络基于真实社会关系(如亲属、朋友、同事等)和现实的物理空间而建立，社交关系简单，而在线社交网络集真实社会关系与虚拟社交关系于一体，跨越了真实的物理空间，在线社交关系庞大复杂，在线社交工具逐渐成为信息传播的新媒介，使人际关系的演变发生着深刻的变革。

微信是腾讯公司在 2011 年推出的一个为智能手机提供即时通讯服务的免费在线社交网络平台。微信可以提供多种功能,包括向好友分享图片、文字,分组聊天、语音,一对多广播消息,视频对讲、位置共享、视频或图片共享、互联网理财及购物、微信支付、信息交流联系、游戏等。同时,用户也可以通过搜索号码、摇一摇、扫二维码、搜索附近的人、漂流瓶等多种方式关注公众平台或者添加好友[1-5] 。微信支持跨通信运营商、跨操作系统平台。截止到 2014 年 6 月，微信已经有了超过 6 亿的用户，用户呈几何级数增长，已成为移动互联网最大的流量入口和亚洲最大即时通讯软件。

随着微信的快速发展，微信成为信息传播的新媒介，人们开始使用微信与家人、朋友、同事联系，分享文字和图片，建立自己的"朋友圈"；也可进行网络营销、购物与支付； 还可以接收新闻和广告信息等，给人们的日常生活、交友和工作带来了很大便捷。

同时，微信在线社交网络快速发展为电子商务的发展带来了机遇，基于微信平台的移动电商—"微商"也随之快速发展，微商是指在微信在线社交平台上，利用微信朋友圈来销售产品进行或提供服务，是一种新出现的销售模式和创业平台。因此本文将研究微信用户关系网络的内部特征，探讨和分析微信用户关系（朋友圈）的拓扑结构和演化模型。

微信在线社交网络本身就是一个大型的复杂网络，有许多学者把复杂网络理论应用到在线社交的研究中：Ebel[6]和Ahn[7]等分别实证了电子邮件和韩国社交网站Cyworld是典型的无标度网络，度分布近似服从幂律分布。罗由平[8]等实证了娱乐名人的微博子网络是小世界无标度网络。Zhang等[9]基于BA模型理论构建了一种增长、择优连接的在线社交网络演化模型，仿真结果表明该在线社交网络度分布服从幂律分布。Xiong[10]等基于博客网络特征构建了一种边和节点同时增长网络模型，该类型的在线社交网络的度分布也服从幂律分布。王亚奇[11]等给出了节点带有吸引度的微博有向网络演化模型，具有无标度特性.度分布指数不仅与反向连接概率有关,而且还取决于节点的吸引度分布。文献[12-15]基于复杂网络理论知识，研究了社交网络上的信息传播演化模型、传播机制及信息推荐机制。文献[16]研究了微信在学术图书馆服务中的应用，文献[17]对大学生的微信接收信息进行了研究,文献[18]基于复杂网络的视角和 Agent 代理，研究微信在线社交网络知识传播演化模型，文献[19]采用复杂网络的视角，实证研究了微信子网络的拓扑结构特征及与之相关的信息传播。文献[20]给出了带有微信功能的物联网框架，从技术的角度给出微信在物联网上的应用框架。用复杂网络理论研究在线社交网络较多，然而采用超网络理论研究在线社交网络论文比较少。研究微信应用的文章较多，然而对微信用户关系的形成和演化机制的研究则很少有人涉及。

本文采用超网络的研究视角，以微信社交关系网络为研究对象，将个人在微信中的朋友关系（朋友圈）看作超边，将个人用户账号看作节点，节点按Poisson过程进入网络，建立了一个节点具有寿命和吸引度的微信用户关系非均齐超网络演化模型，利用Poisson过程理论和连续化方法对模型进行分析，获得超网络稳态平均超度分布的解析表达式，研究结果表明：随着网络规模的增长，老化指数增大，这个动态演化的非均齐超网络的超度分布由幂律分布向指数分布转变。仿真结果和解析结果相吻合。这一研究结果可以为基于微信平台的电子商务（微商）在市场营销策略提供理论上的参考；也可为在线社交网络的传播动力学研究、网络传播控制以及信息的推荐提供某些借鉴意义。

# 2 超网络定义

目前对超网络的研究尚处于起步阶段，关于超网络的概念尚没有统一的定义，目前对

超网络的定义主要分为两类：

(1) 一类超网络的定义是由Denning[21]提出的：基于网络的超网络(Supernetworks)，即网络中的网络NON（Networks of Networks）。

(2) 另一类超网络的定义是Esrada[22-23]提出的基于超图(Hypergraph)的超网络(Hypernetworks)，凡是可以用超图表示的网络就是超网络。文本的研究是基于超图理论的视角。

超图的定义[21][24]如下：

设 $V = \{v_1, v_2, ..., v_n\}$ 是一个有限集，若

(1) $E_i \neq \phi (i=1,2,..,m)$，

(2) 且 $\bigcup_{i=1}^{m} E_i = V$，记 $E^h = \{E_1, E_2, ..., E_m\}$

则称二元关系 $H = (V, E^h)$ 为超图，简记为 $(V, E^h)$ 或 $H$。其中 $V$ 的元素称为超图的节点，$E^h$ 中的元素称为超图的超边。如果两个节点属于同一条超边，则称这两个节点邻接；如果两条超边的交集不空，称之为这两条超边邻接。如图1（a）中，超图 $H = (V, E)$，顶点集 $V = \{v_1, v_2, v_3, v_4, v_5, v_6\}$，超边集 $= \{E_1, E_2, E_3, E_4, E_5\}$，$E_1 = \{v_4, v_2, v_6\}$，$E_2 = \{v_1, v_4\}, E_3 = \{v_1, v_3, v_5\}, E_4 = \{v_3, v_5, v_6\}, E_5 = \{v_2, v_5, v_6\}$，$E_6 = \{v_1, v_6\}$。如果每条超边中的顶点数都相等，则称为均匀超图或一致超图。如果 $|V|$ 和 $|E^h|$ 均有限，则称 $H$ 为有限超图，$H = (V, E^h)$ 退化为图。也可以采用不同方法来描述复杂系统，如图1（a）中的超图可以转化复杂网络理论中的二分图。如图1（b）所示。

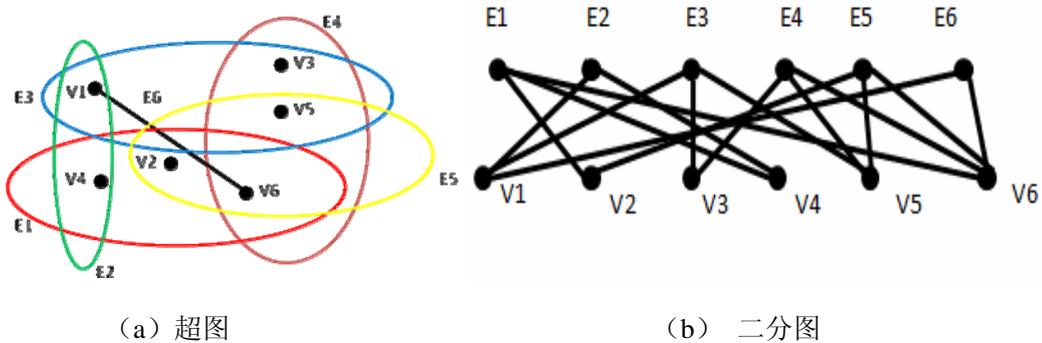

（a）超图　　　　　　　（b）二分图

**图1** 超图和二分图示意图

超网络[25]是超图的推广，假设 $\Omega = \{(V, E^h), |(V, E^h)$ 是有限超图$\}$，$G$ 是从 $T = [0, +\infty)$ 到 $\Omega$ 的映射，对于任意给定的 $t \geq 0$，$G(t) = (V(t), E^h(t))$ 是一个有限超图，指标 $t$ 通常被认为是时间。超网络 $\{G(t)|t \in T\}$ 是指超图的集合。

## 3 微信非均齐超网络演化模型

微信在线社交网络规模庞大，拓扑结构复杂，并且存在安全性问题，直接将微信在线社交网络平台作为实验对象进行研究和分析是十分困难的[7]。因此，本文试图根据微信在线社交网络的特点，对微信网络的拓扑结构进行模型抽象，从而以拓扑模型代替实际的微信在线社交网络，通过拓扑模型认识社交网络的基本特性并进行相关研究。

微信在线社交网具有以下特点：

（1）在微信在线社交网络中，将个人的朋友圈视作超边，朋友圈中的好友看作节点，节点按 Poisson 过程进入网络，在每时每刻，都可能有新用户申请账号加入社交网络，当新用户个体加入微信社交网络后，会选择老用户成为朋友关系，这时，微信在线社交网络中就会形成一个新的朋友圈，也就是形成一条新的超边。每个新个体选择的好友的数量是不一致的，即具有非均齐特点。在微信用户关系超网络中，用户朋友圈中的人数越多，该微信用户越容易被连接到新加入的用户。个人用户所拥有的朋友圈个数为该用户所代表的网络节点的超度。超度越大的用户，拥有的朋友圈越多，越容易与新加入的朋友形成新的超边。微信用户朋友关系超网络示意图如图2所示：

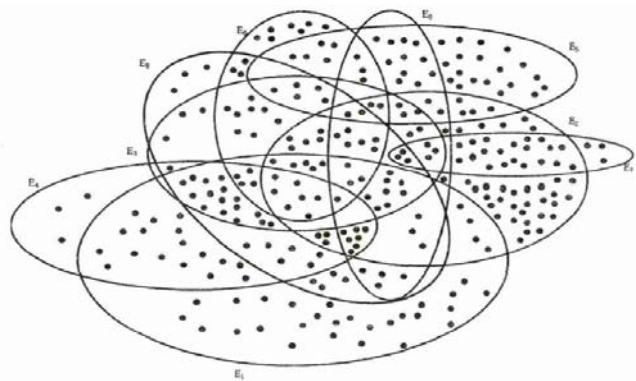

**图2** 微信用户朋友关系超网络示意图 E1，E2..E3 代表朋友圈，节点代表用户。

（2）一些新加入网络不久的节点由于某种经验或声望等某种原因，也可能成为被迅速关注的对象，那么就会出现该节点朋友圈的快速增长，成为微信社交网络中表现活跃的"积

极分子"，节点的这种不同的"活跃性"我们认为：新加入在线社交网络的每一个（或批）个体的拥有的"吸引度"不同，吸引用 $y_0, y_1 \cdots y_n$ 来表示，如图3所示。吸引度可以是均匀分布、指数分布或幂律分布等其中的一种。节点所带的吸引度越大户，越容易与微信在线社交网络中的老节点形成新的超边。

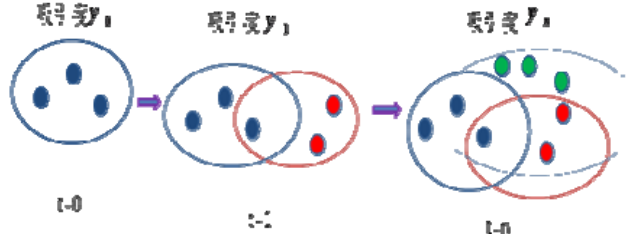

**图3** 带有不同吸引度的节点进入微信在线社交网络示意图

（3） 人们在起初申请一个微信账号加入微信社交网络，我们通常的表现比较积极-活跃度较高，随着时间的推移，我们一般会降低活跃度，使用率不断降低，甚至最终可能不再与账号中的朋友联系。这可能是我们转向其他在线社交应用或者是我们失去"新鲜感"。这种活跃度的降低我们可以看作节点不断的老化。即节点具有寿命。

关于超网络演化模型是近几年的研究热点，Wang 等给出了一种基于超图的超网络动态演化模型，该模型采用增长和优先连接机理逐步生成超网络，每次新增加若干个节点，将这若干个新节点和网络中已有的1 个节点结合生成超边，且每个时间只增加一条超边[26]。胡枫等给出了另一种基于超图的超网络动态演化模型，该模型增长机理与王建伟等模型相对偶，每次新增加1 个节点，将这个新节点和网络中已有的若干个节点结合生成超边，且每个时间只增加一条超边[27]。我们在文献[28]中研究了一类统一模型。然而，这些模型均没有考虑节点吸引力和老化问题。以下我们依据微信网络的特点，建立Poisson连续时间增长的节点老化超网络演化模型，模型描述如下：

1) 初始化：开始于较少的节点数量（$m_0$）的连通超图；

2) 增长：开始于较少的节点数量（$m_0$）的超图，节点到达过程是具有强度为 $\lambda$ 的Poisson过程 $N(t)$，在 $t$ 时刻，当第 $N(t)$ 批新节点进入网络时，从概率密度为 $\rho(\eta)$ 的正整数中抽取 $\eta_{N(t)}$，从概率密度为 $f(y)$ 的非负实数中抽取 $y_{N(t)}$，且 $m_1 = \int \eta \rho(\eta) d\eta < +\infty$， $a = \int yf(y)dy < +\infty$. 第 $N(t)$ 批新节点指定的吸引力为 $y_{N(t)}$，这 $\eta_{N(t)}$ 个新节点与网络中已有的 $m_2$ 个老节点围成一条超边，共围成 $m$ 条超边，且不出现重超边，$mm_2 \leq m_0$；

3) 择优连接： 在选择新节点的连接时，假设新节点连接到节点 $i$ 的概率 $\prod$ 取决于节点 $i$ 的超度数 $k_i$ 和它的年龄，即满足

$$\prod(k_i) = \frac{(k_i + y_i)(t - t_i)^{-1/2}}{\sum_i (k_i + y_i)(t - t_i)^{-1/2}} \tag{1}$$

其中，$t_i$ 是节点 $i$ 进入微信社交网络的时刻，$t - t_i$ 为节点 $i$ 的年龄。

# 4、模型分析

假设 $k(t_i, t)$ 为节点 $i$ 在时刻 $t$ 的超度，则它满足下面的动态方程

$$\frac{\partial k(t_i, t)}{\partial t} = \lambda mm_2 \frac{(k(t_i, t) + y_i)(t - t_i)^{-1/2}}{\sum_j [k(t_j, t) + y_j](t - t_j)^{-1/2}}, \quad k(t, t) = m \tag{2}$$

方程(2) 中 $\lambda$ 表示新节点的到达率。

$$\sum_j (k(t_j, t) + y_j)(t - t_j)^{-\frac{1}{2}} = \int f(y) \int_0^t (k(s, t) + y)(t - s)^{-\frac{1}{2}} \lambda m_1 ds dy$$

方程(2)化为

$$\frac{\partial k(s, t)}{\partial t} = mm_2 \frac{(k(s, t) + y)(1 - s/t)^{-\frac{1}{2}}}{m_1 \int f(y) \int_0^t (k(s, t) + y)(1 - s/t)^{-\frac{1}{2}} ds dy}, \quad k(t, t) = m \tag{3}$$

令 $x = \frac{s}{t}$，$k(s, t) + y \equiv g(x, y)$

那么，方程(3)化为

$$-\frac{s}{t^2} \frac{\partial g}{\partial x} = mm_2 \frac{g(x, y)(1 - x)^{-\frac{1}{2}}}{tm_1 \int f(y) \int_0^1 g(x, y)(1 - x)^{-\frac{1}{2}} dx dy}, \quad g(1, y) = m + y \tag{4}$$

记 $\theta = \int f(y) \int_0^1 g(x,y)(1-x)^{-\frac{1}{2}} dx dy$，则

$$\frac{1}{g(x,y)}\frac{\partial g}{\partial x} = -\frac{mm_2}{\theta m_1 x\sqrt{1-x}}, \quad g(1,y) = m+y \tag{5}$$

对于方程(5)两边从 $\frac{t_i}{t}$ 到1积分，得

$$\ln\frac{g(\frac{t_i}{t},y)}{m+y} = \ln\left(\frac{1+\sqrt{1-\frac{t_i}{t}}}{1-\sqrt{1-\frac{t_i}{t}}}\right)^{\frac{mm_2}{\theta m_1}}$$

所以

$$k(t_i,t) = g(\frac{t_i}{t}) = (m+y)\left(\frac{1+\sqrt{1-\frac{t_i}{t}}}{1-\sqrt{1-\frac{t_i}{t}}}\right)^{\frac{mm_2}{\theta m_1}} - y \tag{6}$$

其中，$\theta$ 是如下特征方程的正解

$$\int_0^1 \left(\frac{1+x}{1-x}\right)^{\frac{mm_2}{\theta m_1}} x dx - \frac{mm_2}{2m_1(m+a)} - \frac{1}{2} = 0 \tag{7}$$

从（6）式，我们有

$$P(k(t_i,t) \geq k) = P\left(t_i \leq \frac{4(m+y)^{\frac{m_1\theta}{mm_2}}(k+y)^{\frac{m_1\theta}{mm_2}}}{((k+y)^{\frac{m_1\theta}{mm_2}} + (m+y)^{\frac{m_1\theta}{mm_2}})^2} t\right) \tag{8}$$

由Poisson过程理论，我们知道节点到达时间 $t_i$ 服从参数为 $i$ 和 $\lambda$ 的Gamma分布 $\Gamma(i,\lambda)$，因此，

$$P\left(t_i \leq \frac{4(m+y)^{\frac{m_1\theta}{mm_2}}(k+y)^{\frac{m_1\theta}{mm_2}}}{((k+y)^{\frac{m_1\theta}{mm_2}} + (m+y)^{\frac{m_1\theta}{mm_2}})^2} t\right)$$

$$= 1 - e^{-\frac{4(m+y)^{\frac{m_1\theta}{mm_2}}(k+y)^{\frac{m_1\theta}{mm_2}}}{((k+y)^{\frac{m_1\theta}{mm_2}} + (m+y)^{\frac{m_1\theta}{mm_2}})^2}\lambda t} \sum_{l=0}^{i-1} \frac{1}{l!}\left(\frac{4(m+y)^{\frac{m_1\theta}{mm_2}}(k+y)^{\frac{m_1\theta}{mm_2}}}{((k+y)^{\frac{m_1\theta}{mm_2}} + (m+y)^{\frac{m_1\theta}{mm_2}})^2}\lambda t\right)^l \tag{9}$$

由此，我们可以获得连续增长微信社交网络模型的稳态平均超度分布为

$$P(k) \approx \frac{4m_1\theta}{mm_2} \int \frac{(m+y)^{\frac{m_1\theta}{mm_2}}(k+y)^{\frac{m_1\theta}{mm_2}-1}[(k+y)^{\frac{m_1\theta}{mm_2}} - (m+y)^{\frac{m_1\theta}{mm_2}}]}{[(k+y)^{\frac{m_1\theta}{mm_2}} + (m+y)^{\frac{m_1\theta}{mm_2}}]^3} f(y)dy$$

(10)

其中，$\theta$ 是如下特征方程(7)的正解.

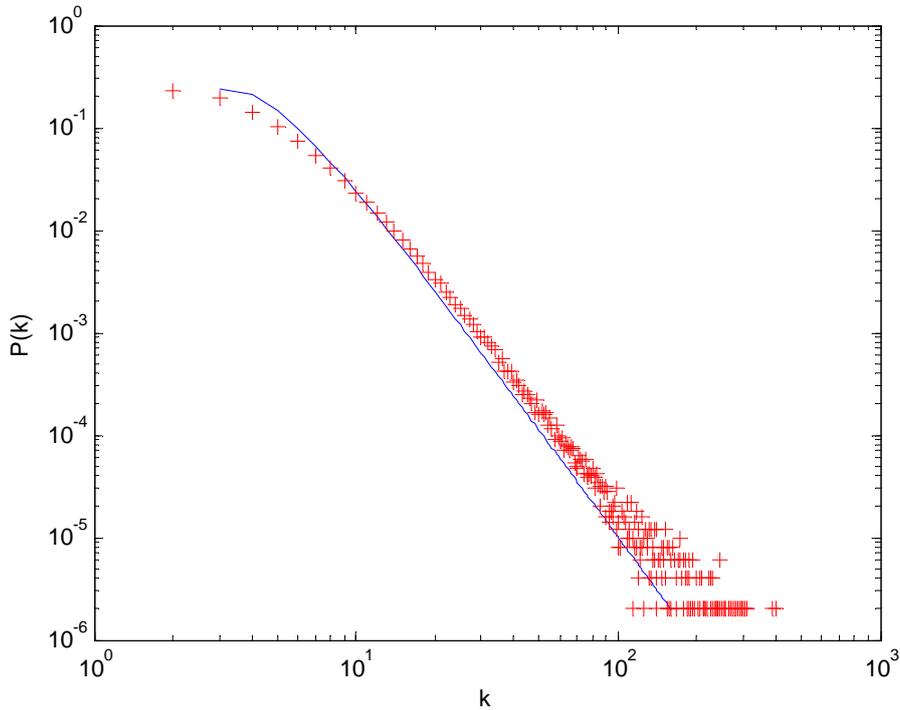

**图4** 微信社交网络模型模拟图。 +表示模拟度分布，直线表示理论公式(10)预测的理论曲线。网络规模100000，初始网络20个节点，新节点在1~5的整数中随机抽取，$m=2$，$m_2=6$，节点吸引力为 $y_{N(t)}$ 服从[0，1]上的均匀分布。

# 5 老化指数普适性模型

上述分析和数学解析假设 Poisson 连续时间增长社交网络模型老化指数为 $\alpha=1/2$，不失一般性，下面我们建立老化指数普适性的模型并利用计算机仿真研究老化指数对于微信

社交网络超度分布的影响。

模型：增长与 Poisson 连续时间增长的节点老化超网络演化模型相同；在选择新节点的连接时，假设新节点连接到节点 $i$ 的概率 $\prod$ 取决于节点 $i$ 的度数 $k_i$ 和 $i$ 的年龄 $t-t_i$，即满足

$$\prod(k_i) = \frac{(k_i + y_i)(t-t_i)^{-\alpha}}{\sum_i (k_i + y_i)(t-t_i)^{-\alpha}} \tag{11}$$

其中，$t_i$ 是节点 $i$ 进入网络的时刻，$\alpha$ 是一个非负常数。

这个模型具有一定的普适性，文献[29]中的模型也是这个模型的特殊情形。这个模型的几个特例仿真参见图 5，可见随着 $\alpha$ 的变大分布指数不断增大，当 $\alpha=1$ 时，超度分布接近于指数分布。这也是可以理解的，因为节点的超度不会超过 $\lambda t$，当 $\alpha=1$ 时，择优条件(11)就接近于随机选择。

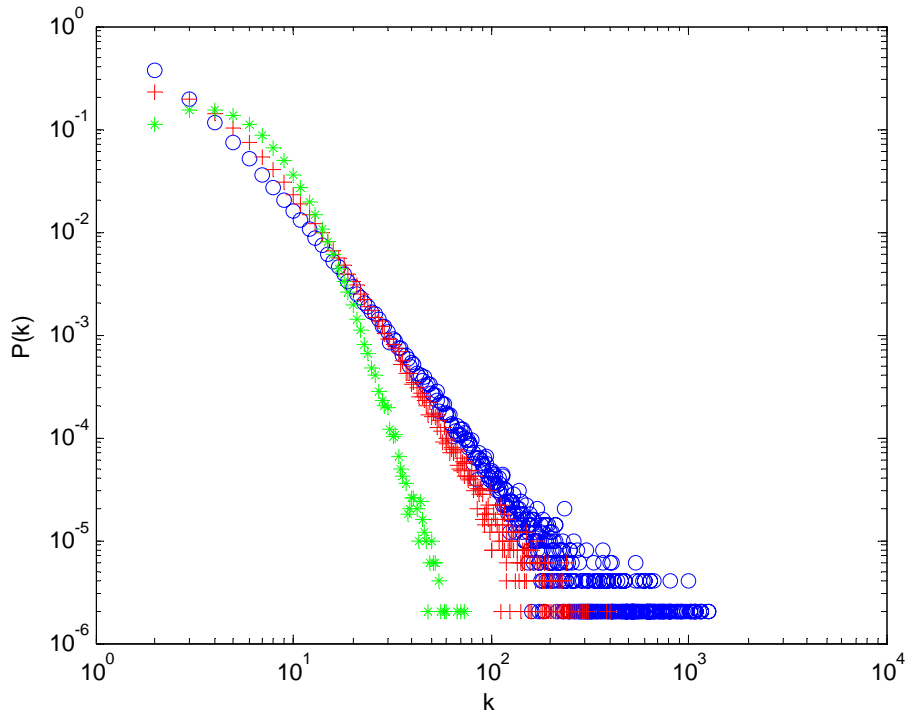

图5 带有老化指数普适性的微信社交网络模拟图。O 表示 $\alpha=0.1$ 模拟度分布，+ 表示 $\alpha=0.5$ 模拟度分布，*表示 $\alpha=0.9$ 模拟度分布。网络规模100000，初始网络20个节点，新节点在1~5的整数中随机抽取，$m=2$，$m_2=6$，节点吸引力为 $y_{N(t)}$ 服从[0，1]上的均匀分布。

# 6 结论

在微信用户关系超网络演化模型研究中发现，随着老化指数的增大，超度分布由幂律分布向指数分布转化。这一研究发现：可以为微商用户营销决策提供如下的管理启示：

(1)当微信超网络的超度分布具有幂律分布时，这时的超网络具有"马太效应"，即有"富者越富"效应。即当"微商"的朋友圈形成的超网络具有"幂律"分布，"潜在的客户群（或朋友圈）"与微商连接时，容易形成"择优选择"，此时的微信（微商）用户关系超网络可以得到更多的市场份额和潜在用户，容易形成规模，取得最优的经济效益。

（2）如果节点的老化加快，微信超网络的超度分布由幂律分布转化为指数分布时，"潜在的客户群（或朋友圈）与"微商"的容易连接由"择优选择"向"随机选择"转化，这时会导致潜在的客户群比较分散，不容易形成规模效益。

为了避免整个微信商圈网络中的节点老化现象，避免节点的兴趣度降低和节点退出，防止微信在线超网络由幂律分布向指数分布转化， 因此"微商"在进行市场营销时，应及时地维护良好的客户互动关系，对一些不太活跃的用户和一段时间没有联系的客户，利用节假日和促销周进行问候或推送产品信息，或者定期浏览不活跃用户分享的信息，及时掌握不活跃用户的现状，不定期的建立联系。"微商"也要通过推送一些引起朋友圈兴趣的内容，保持"节点"一直处于活跃状态，积极避免客户节点的老化和吸引度下降的现象，维持微信超网络处于幂律分布状态以取得最优的经济效益。

在线社交关系复杂繁多，其演化过程非常复杂，只有理解在线社交网络上的演化过程，建立符合实际的演化模型，我们才能够有依据地进行社交网络的传播动力学研究、网络传播控制以及信息的推荐等的研究。